\documentclass[aps,prd,twocolumn,groupedaddress,eqsecnum,amssymb,showpacs,
nofootinbib]{revtex4}
\usepackage{graphicx,mathptm,bm,times,epsfig}

\newcommand{\be}{\begin{equation}}
\newcommand{\ee}{\end{equation}}
\newcommand{\bea}{\begin{eqnarray}}
\newcommand{\eea}{\end{eqnarray}}

\begin{document}

\title{Cosmological Avatars of the Landscape II:\\  CMB and LSS Signatures }

\author{R.~Holman}
\email[]{rh4a@andrew.cmu.edu}
\affiliation{Department of Physics, Carnegie Mellon University, Pittsburgh PA 15213, USA}
\author{ L.~Mersini-Houghton}
\email[]{mersini@physics.unc.edu}
\affiliation{Department of Physics and Astrononmy, UNC-Chapel Hill, NC, 27599-3255, USA}
\author{ T.~Takahashi}
\email[]{tomot@cc.saga-u.ac.jp}
\affiliation{Department of Physics, Saga University, Saga 840-8502, Japan}

\date{\today}

\begin{abstract}
This is the second paper in the series that confronts predictions of
a model of the landscape with cosmological observations. We show here how the
modifications of the Friedmann equation due to the decohering effects
of long wavelength modes on the wavefunction of the Universe defined
on the landscape leave unique signatures on the CMB spectra and
large scale structure (LSS). We show that the effect of the string
corrections is to suppress $\sigma_8$ and the CMB $TT$ spectrum at
large angles, thereby bringing WMAP and SDSS data
for $\sigma_8$ into agreement. We find interesting features imprinted on the matter
power spectrum $P(k)$: power is suppressed at large scales
indicating the possibility of primordial voids competing with the ISW
effect. Furthermore, power is enhanced at structure and substructure scales,
$k\simeq 10^{-2-0} h~{\rm Mpc}^{-1}$. 
Our smoking gun for discriminating this proposal
from others with similar CMB and LSS predictions come from
correlations between cosmic shear and temperature anisotropies, which
here indicate a noninflationary channel of contribution to LSS, with
unique ringing features of nonlocal entanglement displayed at
structure and substructure scales.
\end{abstract}


\maketitle

\section{Introduction}
\label{sec:intro}
While finding a dynamical reason for why certain initial conditions for the Universe were preferred over others would be a major step forward in our understanding of Nature, such a proposal must also be falsifiable. In previous work \cite{richlaura1,richlaura2}, we have argued that the quantum dynamics of gravity acting on the string landscape, treated as the configuration space of initial conditions \cite{land1,land2}, could in fact show why the initial conditions for high scale inflation are natural. 

In the previous paper of this series (hereafter referred to as paper
$I$) \cite{avatars1}, we showed how tracing out long wavelength modes of metric and matter field fluctuations could decohere the
wavefunction of the Universe, defined on this configuration space. This results in both inducing a
highly nonlocal entanglement between our horizon patch with others as
well as giving rise to corrections to the Friedmann equation. We then used these cosmological
effects to bracket the scale of SUSY breaking $b$ which appears as one of the
parameters determining the size of the nonlocality of
entanglement. The SUSY breaking bounds were derived by requiring that
constraints on the flatness of the modified inflaton potential and the
ammount of density perturbations be satisfied when the quantum gravity
corrections are taken into account.

The backreaction effects that modify the Friedmann equations during inflation (essentially
correcting the potential) will also modify the power
spectrum. These modifications affect a number of
cosmological observables, such as the CMB temperature anisortopy (the
TT spectrum), large scale structure (LSS), the creation and statistics
of voids as well as running of the scalar spectral index $n_s$. We
discuss these in detail in Sec.~\ref{sec:sig}. In particular, we find
a natural explanation for the disagreeement between SDSS and WMAP data
on the low value of $\sigma_8$ and for the observed low $l$ anomaly in
the CMB (temperature anisotropies) $TT-$ spectrum observed both by
COBE \cite{cobe} as well as by WMAP \cite{wmap3}. We then turn to a
discussion of how the backreaction effects may contribute with a
negative density contrast that gives rise to the creation of
primordial voids in Sec.~\ref{sec:sig}. We also show how the effects we find here can be distinguished from other effects predicted by other models of quantum gravity. The main tool we use to do this is the cross-correlation $C_{l ,TL}$ of temperature (T) with the lensing potential of cosmic shear (L).

We collect all of our results and discuss their regimes of validity in
Sec.~\ref{sec:smoke}, and then conclude.

\section{Quantum Effects of the Landscape}
\label{sec:quant}

For completeness, we recall some of the basic relations found in
paper $I$. As in paper $I$, we follow Ref.~\cite{katielaura} and use the following
inflaton potential
\be
\label{eq:infpot} V(\phi) = V_0 \exp\left(-\lambda
\frac{\phi}{M_{\rm P}}\right). \label{eq:V_inf} 
\ee 
While this potential can appear in supergravity based models of
inflation, we use it more for illustrative purposes, due to its
simplicity. Our results will hold generically for any inflaton
potential.

After including the backreaction due to the long wavelength modes
being traced out, the effective Friedmann equation was shown to
be \cite{avatars1}: 
\be
\label{eq:modfried} H^2 = \frac{1}{3 M_{\rm P}^2}
\left[V(\phi)+\frac{1}{2} \left(\frac{V(\phi)}{3 M_{\rm
P}^2}\right)^2 F(b,V)\right]\equiv \frac{V_{\rm eff}}{3 M_{\rm P}^2}
\ee
where
\bea \label{eq:corrfactor}
F(b,V) 
&=& 
\frac{3}{2} \left(
2+\frac{m^2M^2_{\rm P}}{V}\right)
\log \left( \frac{b^2 M_{\rm P}^2}{V}\right)\nonumber \\
&-&
\frac{1}{2} 
\left(1+\frac{m^2}{b^2}\right) \exp\left(-3\frac{b^2
M_{\rm P}^2}{V}\right). 
\eea
We have taken $8\pi G_N = M_{\rm P}^{-2}$ and the scale of the
nonlocal entanglement is given by the interference length $L_1$,
derived in \cite{avatars1}.
\be
L_1^2 =
\frac{a}{H}  \left[ \left(\frac{m^2}{3 H}+H\right)\ln\frac{b}{H}-\frac{m^2 H}{6}\left(\frac{1}{b^2}-\frac{1}{H^2}\right) \right].
\label{eq:L1}
\ee

The primordial power spectrum is estimated from 
\be P_\mathcal{R} =
\frac{1}{75 \pi^2 M_{\rm P}^2 } \frac{V_{\rm eff}^3}{V_{\rm eff}^{'2}}. 
\ee 
For the initial potential, given
in Eq.~(\ref{eq:infpot}), we have \be P_\mathcal{R}^0 =
\frac{1}{75\pi^2 M_{\rm P}^2 } \frac{V_0}{ \lambda^2 M_{\rm P}^4} \ee
The scalar spectral index before modifications is given by $n_s^{0} -1
= -\lambda^2$. Modifications in the Friedmann equation result in a
running of the spectral index $n_s = n_s^{0} +\delta n_s$, as we
describe below.

The solution for the inflaton field in the presence of the corrected
potential is given by: 
\bea \phi &=& \lambda M_{\rm pl} \left[ 1 +
  \frac{1}{2} \frac{1}{ 3M_{\rm pl}^2} \left(\frac{V_0}{3 M_{\rm
      pl}^2} \right) \right. \nonumber \\ && \left. \times \left\{ 3
  \left( 2 + \frac{m^2 M_{\rm pl}^2}{V_0} \right) \log \left( b \sqrt{
    \frac{3M_{\rm pl} }{V_0}} \right) \right.  \right. \nonumber \\ &&
  \left. \left. -\frac{1}{2} \left( 1 + \frac{m^2}{b^2} \right) e^{-
    3M_{\rm pl}^2 b^2 /V_0} \right\} \right]^{-1} \log \left(
\frac{k}{k_{\rm ref}} \right) \nonumber \\
\label{eq:fieldsol}
\eea 
where $k_{\rm ref} \simeq (4000 ~{\rm Mpc})^{-1}$.

Define $3M_p^{2}\slash F(b,V) \equiv \sigma(b,\phi)$ and denote the
energy correction $V^2 \slash \sigma = f(b,V)$. The modified Friedmann
equation can then be written as 
\be
\label{eq:friedshorthand}
3M_p^{2} H^2 = V + f(b,V)
\ee
An important fact is that $f(b,V)$ is a negative function, so that the
new Friedmann equation only makes sense in the regime where 
$ V + f(b,V)>0$.

We are now prepared to examine the effects of this modification on
large scale structure and the CMB.

\section{Signatures of the Landscape on CMB and LSS}
\label{sec:sig}

Armed with our modified Friedmann equation and power spectrum, we are now ready to compute the corrections to the CMB power spectrum and to Large Scale Structure (LSS). 

\begin{figure}[!htbp]
\begin{center}
\raggedleft \centerline{\epsfxsize=3.5in \epsfbox{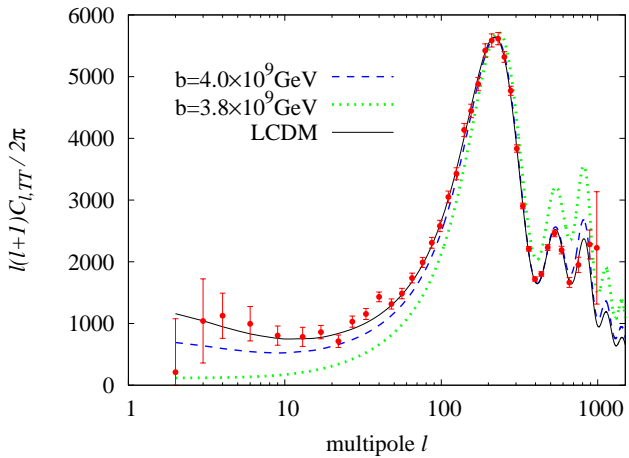}}
\caption{CMB TT power spectra for the cases with $b=4.0 \times 10^9$
GeV (dash-line) and $3.8 \times 10^9$ GeV (dot-line).  For reference, the spectrum for the
$\Lambda$CDM case (solid-line) and the data from WMAP3 are also plotted.} 
\label{fig:fig1}
\end{center}
\end{figure}

In Fig.~\ref{fig:fig1} we show the results obtained using CMBFAST \cite{cmbfast} with
the modification in the primordial spectrum of Eq.~(\ref{eq:modfried})
for $b\simeq 10^{-9} M_{\rm P}$.  For comparison, we also plot the case with 
$\Lambda$CDM and data from WMAP3 \cite{wmap3}.
For the plot, we fixed the parameters for the inflaton potential as $\lambda=0.1$ 
and $V_0 = 8.0 \times 10^{-8} \ M_{\rm P}^4$ and $1.0 \times 10^{-8}\ M_{\rm P}^4$
for $b=4.0 \times 10^8$ GeV and $3.8 \times 10^9$ GeV 
respectively. The cosmological parameters are taken to be 
$\Omega_mh^2 = 0.12, \Omega_bh^2 = 0.023, h= 79,  \tau = 0.091$ 
and $n_s = 0.99$ for LCDM models. For the cases with our model,
we varied them slightly to fit the spectrum to WMAP3 data around the acoustic peaks. 

It can be seen from Fig.~\ref{fig:fig1} that power at low $l$ is suppressed compared to the concordance LCDM model.

Consistency with the Einstein equations in deriving the primordial
spectrum requires that we include the pressure corrections
corresponding to the energy modification Eq.~(\ref{eq:modfried}). This
correction directly contributes to the velocity of the inflaton
field $\dot\phi$ and therefore to the running of the scalar index
$n_s$ as discuss further below.  The pressure contribution from the
correction term for the inflaton with energy density $\rho \approx V$
is 
\be 
p_f = (\rho+p)\frac{df(b,V)}{d\rho} -f(b,V)
\label{eq:pressure}
\ee

Including this modification into the expression for the primordial
spectrum and replacing the field solutions, Eq.~(\ref{eq:fieldsol}), in
$P_{\mathcal R}$ in order to obtain $P_{\mathcal R}$ as a function of
the wavenumber $k$ we get 
\be 
P_{\mathcal R} 
= 
\frac{1}{75\pi^2 M_{\rm  P}^2 } 
\frac{V_0(1 + V_{0}/2\sigma)^3}{ \lambda^2 M_{\rm P}^4}\ g(b,\phi)
\simeq 
\left(\frac{k}{k_{\rm ref}}\right)^{n_s  -1},
\label{eq:modprim}
\ee
where
\begin{equation}
g(b,\phi)=\left(1+4\pi\left(\frac{f(b,V) +p_f}{M_P^2 \dot{H}}\right)\right)^{-2}\simeq \left(\frac{k}{k_{\rm ref}}\right)^{\delta n_{s}}
\label{eq:correctns}
\end{equation}
For the parameter choice described above the maximum power is roughly
on scale $180$ Mpc. The modified $P_{\mathcal R}$ is an increasing
function of $k$ for small scales and a decreasing function at large
scales. This gives rise to a suppresion of power on quadrupole scales
and a running of the spectral index $n_s$.  The spectrum tilts a
little to the blue on large scales and has a slight running towards
red in small scales for some parameter sets.
This feature can be seen in Fig.~\ref{fig:fig1}, in particular, 
 for the case with $b= 4.0 \times 10^{9}$
GeV. Now let us look at these effects in more detail.

\subsection{Suppressed $\sigma_8$}

As seen from the figure, the suppresion around the quadrupole results
in a $30 \%$ decrease on the rms amplitude $\sigma_8$. This is a
direct result of the modification of the Newtonian
potential $\delta\Phi$ by the {\it negative} modification term
$f(b,V)$ on quadrupole scales. From the Poisson equation the Newtonian
potential is given by $-2\slash 3 (k\slash aH)^2 \Phi = 4\pi G_n
\rho$. The correction to the gravitational potential is of second order
and comes from the addition of the non-Gaussian perturbation $f(b,V)$
to the adiabatic Gaussian perturbations $V$: $\nabla^2\delta\Phi =4\pi
G_N f(b,V)$, where
\be
\label{eq:modepot}
\Phi =
\Phi^0 + \delta \Phi 
\simeq 
\Phi^{0} \left[ 1 + \frac{f(b, V)} {\rho} \left( \frac{r}{L_1(k,b)} \right)^2 \right].
\ee
The largest contribution to the Newtonian potential from the
modification in the primordial spectrum, Eqs.~(\ref{eq:modprim}) and 
(\ref{eq:correctns}), is at $k\simeq 1$ corresponding to present day scales of $\simeq 10^{-4}$ Mpc, i.e at the quadrupole
level. The induced correction to the quadrupole, calculated in $I$ to
be 
\bea
\label{eq:quadrupole}
&(\nabla T / T)_{\rm quad}& 
\approx 
r_H^{2}\nabla^2 \delta\phi \nonumber \\ 
&=(c k_{1} / H_0 )^2\delta\phi&\approx 0.5 (r_H/L_{1})^2
(\delta\rho / \rho)_{1}, 
\eea 
originates from the inhomogeneities
superimposed on the background potential at scales $L_1$ which is at
least $100$ to $1000$ times the size of the horizon. Note that
$\delta_f\equiv (\delta\rho /\rho)_1\simeq f(b,V) / V$ is
negative. Therefore the corrected potential contributes a
negative density contrast $\delta _f$ which is  superimposed on the
positive gaussian density perturbations. {\it This results in a
lowering of the amplitude $\sigma_8$ by around $20-30 \% $} . It is
very interesting that due to the tight bounds on $b$ found in $I$, the
amplitude suppression is forced to be
$20-30 \%$ less that the amplitude in a $LCDM$ cosmology. This then brings
the WMAP results  for $\sigma_8 \simeq 1.1$ in agreement
with the 2DF and SDSS findings \cite{SDSS} $\sigma_{8} \simeq
0.8$. 

Another interesting fact that we will comment on further below is that the negative density
correction results in a void being generated around $k\simeq
1$. 

\subsection{Running of the Scalar Spectral Index, $\delta n_s$}

The running of the spectral index $\delta n_s =n_s - n_s^{0}$ can be
readily estimated from the Eqs.~(\ref{eq:modprim}) and (\ref{eq:correctns})
above.  Since the modified spectrum is $P_{\mathcal R}\simeq
(k/k_{\rm ref})^{n_s -1}$ and the unmodified power spectrum $P^0_{\mathcal
  R}\simeq (k/k_{\rm ref})^{n_s^{0} -1}$ it follows that 
\be 
\delta n_s 
= 
\frac{d \ln g(b,\phi)}{d\ln k} 
= 
\frac{\dot\phi}{H} \left(\frac{d \ln  g(b,V)}{d \phi} \right).
\label{eq:running}
\ee
 
The running of the spectral index $\delta n_s(k)$ as a function of $k$
can now be estimated by replacing $\dot\phi \approx -V'_{\rm eff}/ 3H$ and
using the solution for $\phi$ in Eq.~(\ref{eq:fieldsol}). The expression
is very long and not particularly illuminating so instead of writing it here
we can just describe the relevant features. The maximum running is at
low wavenumbers, the result of which is a suppression of the
$TT$-spectrum at low multipoles, Fig.~\ref{fig:fig1}. The reason for
this is that the correction term in the Friedman equation has its
maximum effect at the onset of inflation when $V\simeq \sigma(b,V)$,
and therefore the inflaton slow-roll is disturbed near the scales
corresponding to the first few e-foldings. This results in suppressed
power at these scales. As $V$ decreases during inflation the
correction term is redhifted away. The overall running tends towards the 
blue at low multipoles and towards the red at higher multipoles.

Unlike the models in Refs.~\cite{katielaura,Contaldi} the correction
terms that modify the Friedmann equation, and therefore the energy and
length cutoff scales $\sigma(b,V, L_1(b,V))$ in our model are dynamical.
By this we mean that their scale of influence changes with $k$ since
due the interlocking of $b$ and $V$ contained in $\sigma(b,V)$, the
ratio $b^2\slash V(\phi)$ changes as $\phi$ rolls down its potential. This
results in three different regimes for the slow-roll of inflaton and
generation of perturbations. At the onset of inflation, the two energy
terms in $H^2$ are comparable thus the amplitude of perturbations and
$TT$ power is suppressed around the quadrupole. The dominant
contribution in the interference length comes from the exponential
term, which as we mentioned before, arises from tunneling to and from
other patches.

When $k_{18}$\footnote{
Here the notation $k_{18}$ means that the scale which exits the horizon 
$18$ e-folds after the onset of inflation.
}  leaves the horizon, the inflaton potential $V\slash M_P^2$ becomes
comparable to $b^2$  and thus the exponential in the correction term
dominates over the logarithmic term. This results in a discontinuity
of $\delta n_s$ around $k_{20}$ which suppresses the perturbations of
$k\simeq 18-20$, corresponding to the scale $140-180$ Mpc today. This
discontinuity is important for structure formation features as
discussed below.

A second discontinuity occurs around $k_{40}$ towards the end of
inflation where $V$ has dropped so much that the exponential term in
$\sigma(b,V)$ is nearly zero and the energy correction is dominated by
the log term. Furthermore, as $V$ drops below $b^2$, the log term and
$f(b,V)$ change sign from negative to positive energy corrections,
which creates a jump disconinuity at $k\simeq 40$ for $n_s$ and thus
another suppresed perturbation around a scale which at present is deep
into the scales relevant for substructure formation.

\subsection{LSS, Voids and the Axis of Evil}
\label{subsec:lss}

 We described above the effect of the dynamic modification term on the
 background potential. Until $k=40$, this modification is negative and the correction to $\Phi$ lowers structure there. Thus we
 have a perturbation channel originating from the entanglement
 modification term which produces inhomogeneous, nongaussian and
 nonadiabatic anisotropies ${\delta\Phi}\slash{\Phi^0}$ that are superimposed on the gaussian
 perturbations generated by the inflaton potential $V$ ({\it c.f.} Eq.~\ref{eq:modepot}.)
 
The present day Newtonian background potential can be estimated
from Eq.~(\ref{eq:modepot}) through the expression $\Phi(r,z) =
(1+z)G(z)\Phi(r,0)$ where $G(z)$ is the growth factor of structure and
$r,z$ are the physical comoving distances and redshifts
respectively. It should be noted that the correction term here is
scale dependent since it involves a coupling of the $b,V$ channels for
the highly nonlocal entanglement with size $L_1 \gg r_H$.  Since most
of the matter in our universe is made of dark matter we can view the
above modification to the background Newtonian potential as a
perturbation on dark matter, denoted by $(\delta\rho / \rho)_1$ in the
previous section, but of arising due to quantum entanglement, as
opposed to having an inflationary origin.

\begin{figure}[!htbp]
\begin{center}
\raggedleft \centerline{\epsfxsize=3.5in
  \epsfbox{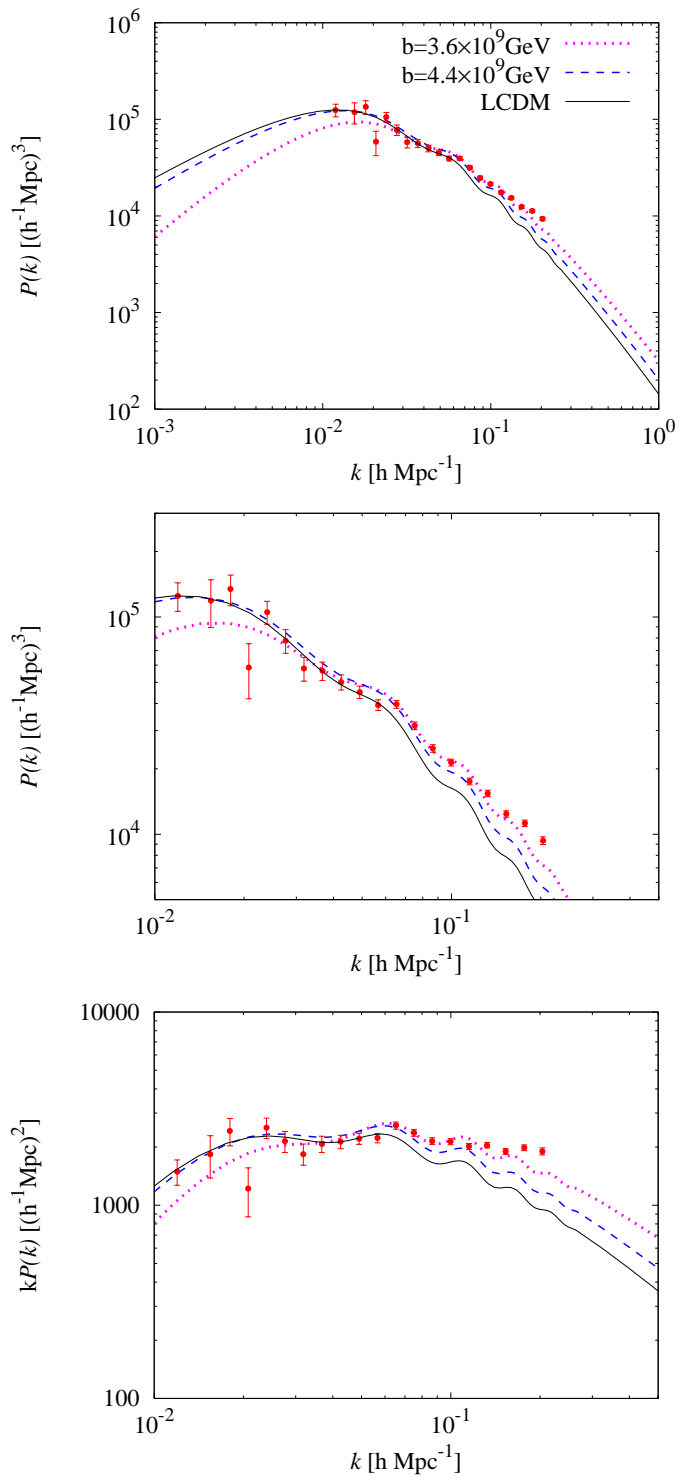}} 
  \caption{
  Matter power spectra for the same values of $b$ as Fig.1, described in Section III. 
  For comparison, the case with $\Lambda$CDM (solid line) and the data 
  from SDSS are also plotted. Top panel shows $P(k)$ for several orders of $k$,
  middle panel is the same as the top panel except the range of $k$ to look 
  at the scales seen by SDSS and bottom panel shows $k P(k)$.
  }
  \label{fig:fig2}
\end{center}
\end{figure}

In Fig.~\ref{fig:fig2}, we plot the matter power spectrum 
calculated with CMBFAST 
with the modifications  in our model. 
For comparison we also plot the results of the LCDM model as well as the SDSS data \cite{SDSS}.
Here the parameters for the inflaton potential are taken 
as $\lambda=0.1$ and $V_0 = 8.0 \times 10^{-8}\ M_{\rm P}^4$.
We varied the overall normalization slightly to fit the data 
by using the uncertainty of the bias. 

As can be seen from Fig.~\ref{fig:fig2} for
the matter power spectrum $P(k)$,  the negative contribution
from the modification of the energy density during inflation,
which takes its maximum value at large scales results in less structure
at very large scales in our model compared to the concordance
cosmology $LCDM$. This is due to the negative correction in
Eq.~(\ref{eq:modepot}) which gives rise to a negative density contrast
denoted above by $\delta_f <0$ being superimposed on the positive
background potential thereby resulting in damped structure formation and the
creation of primordial voids at those large scales.
We discussed above how the correction to the quadrupole amplitude 
which damps power on these scales also produces the first void at around
 $k\simeq 1$, when evolved to present times, resulting also in a compensation of the
ISW effect of dark energy. This void is unaffected by the nonlinear
physics of the subsequent structure growth since it is located at low
multipoles where the potential is nearly constant.  According to Inoue and 
Silk \cite{inouesilk}, a void-induced quadrupole of any size can explain
the two cold spots observed by SDSS\cite{SDSS} in the sky at $(\Delta T/T)_{\rm quad}
<0$ and the alignment of quadrupole with octupole along the preferred
axis which here is the direction that joins the two voids at 60
degrees on the sky, namely the axis of evil. We refer the reader to
Ref.~\cite{inouesilk} for the details of this calculation.

We expect the second void to form at the first discontinuity in the
kinetic term $\dot\phi, V'_{\rm eff}$ corresponding to scales $k=20$
or about $200$ Mpc at redshifts about $z\sim 1$. This void would
correspond to an angle less than a degree in the sky today, which may
be related to the unusual cold spot observed in the southern
hemisphere, as discussed in Ref.~\cite{inouesilk}. However, a deeper understanding of this point requires dealing with the nonlinear growth of structure at these scales. It is unclear to us whether these structures will persist after nonlinear evolution to the present epoch.

Although the second discontinuity would suppress the background
perturbations at around $k=40$, it would not produce voids now
the superimposed nonadiabiatic perturbation from the modification term
has a positive density contrast, which serves to enhance structure
formation rather than suppress it (Fig.~\ref{fig:fig2}) .

Fig.~\ref{fig:fig2}, 
confirms that power is indeed suppressed compared to LCDM at large scales. However,
unlike in the LCDM situation, it is enhanced at structure and substructure scales. 
As far as the linear theory is concerned, 
$P(k)$ in our model is in closer agreement with the SDSS data than the conventional $LCDM$ model.
For $k < 0.09\ h /{\rm Mpc}$, nonlinear evolution can alter the matter 
power spectrum. Thus our model may have implications for the nonlinear evolution of structure.
We would like to emphasize here that we have no room to change 
the parameters $b,V$ in
order to fit the data shown in Fig.~\ref{fig:fig2} since the
effect is derived from a highly nontrivial and nonlocal term that
couples $b,V$. 

\subsection{Cosmic Shear and its Correlation with Temperature Anisotropies, $C_{l,TL}$}
\label{sub:shear}

How can we distinguish our model from other models that might predict
suppression of power at low $l$? This crucial issue was addressed in
\cite{steenlaura} with the result that the best tool to use to discriminate between models
is the prediction for the cross-correlation between the  $TT$ and the cosmic shear $LL$
lensing potential which traces out the large scale structure of the
universe. Doing this gives us a direct comparison between the sources that seed both types of perturbations.The standard picture of the generation of fluctuations in inflationary cosmologies provide only one source,
the primordial spectrum, for seeding both CMB and LSS and therefore
predicts a correlation of order one on large scales. On the other hand, models
with input from quantum gravity/string theory will likely provide
additional sources for perturbations or gravitational potential 
independent of the primordial spectrum since they could contain more degrees of
freedom. Therefore the $TL$ correlation would reflect this input of
new physical degrees of freedom by predicting a correlation different
from one and with unique features on it which are used to identify the
model.

\begin{figure}[!htbp]
\begin{center}
\raggedleft \centerline{\epsfxsize=3.5in
  \epsfbox{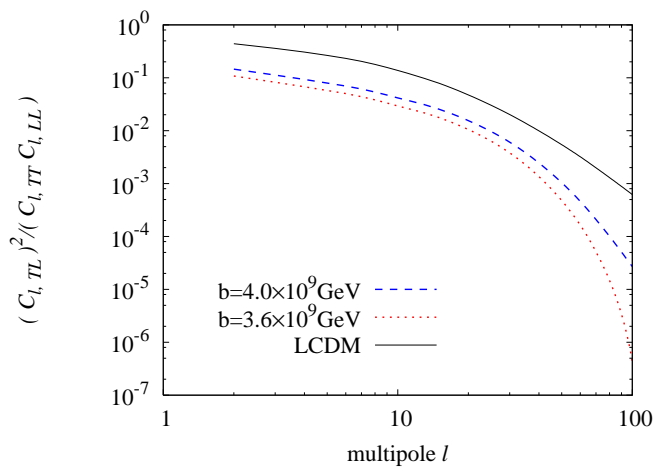}} 
  \caption{Cross correlation between
lensing and temperature are plotted. We assumed $b= 4.0 \times 10^9$ (dash-line)
GeV and $3.6 \times 10^9$ GeV (dot-line) in this figure. For comparison, the case with $\Lambda$CDM (solid-line) is also plotted.} 
\label{fig:fig3}
\end{center}
\end{figure}

\begin{figure}[!htbp]
\begin{center}
\raggedleft \centerline{\epsfxsize=3.5in
  \epsfbox{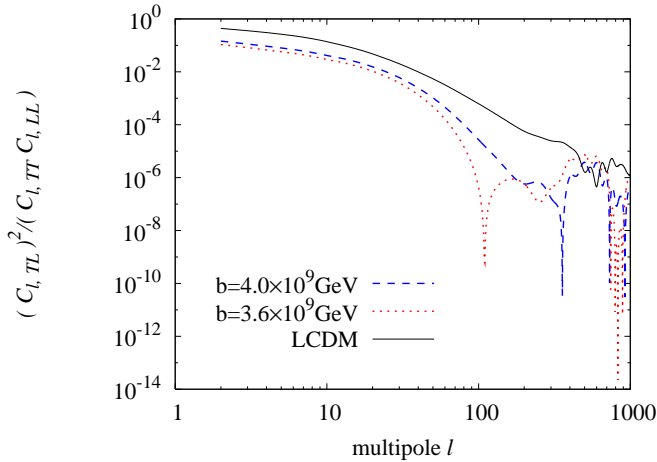}} 
  \caption{The same as Fig.\ref{fig:fig3} except 
we plotted the larger range of multipoles here.} 
    \label{fig:fig4}
\end{center}
\end{figure}

We follow the approach of \cite{steenlaura} in calculating the
cross-correlation fluctuations $C_l^{TL}$ between shear and
temperature. Fig.~\ref{fig:fig3} shows our results for this
cross-correlation. 
In the figure, we take $b=4.0 \times 10^{9}$ and $3.6 \times 10^{9}$ GeV 
as an example. Other parameters are taken as 
$\lambda=0.1$ and $V_0 = 1.2 \times 10^{-8} M_{\rm P}^4$.
The fact that our modification of the Friedmann
equation comes as it does, {\it i.e.} from entanglement effects
between our horizon patch and others gives rise to some interesting
and potentially unique features which we display in
Figs.~\ref{fig:fig3} and \ref{fig:fig4}
 
At large scales (low multipoles) we can see that the cross correlation
of our model has essentially the same shape what would come from a  LCDM model but with a suppressed amplitude. The reason for this is that at those scales near
the onset of inflation, the correction cutoff scale $\sigma(b, k)$ (which is obtained by substituting Eq.~\ref{eq:fieldsol} into the definition of $\sigma(b,\phi)$) is nearly
a constant and the overall effect of its negative energy contribution
is to slightly suppress $TT$ spectrum, Fig.~\ref{fig:fig3}. The most
interesting and distinct features can be found at shorter scales $180< l
<700$ as shown in Fig.~\ref{fig:fig4}. We see that around $l\simeq
200$ the correlation $C_l^{TL}$ dips by about three orders of
magnitude, with a ringing effect at structure formation scales and
another dip in power at substructure scale, with ringing in
between. The damping of this correlation means that much of the
contribution to LSS is not coming from the primordial inflationary
channel at those scales but from the correction term to the Newtonian
potential given by $\sigma(b,k)$. We believe that these features
correspond to the discontinuity found in the spectral index at around
$k=20, k=40$ which arises from the changing in the dynamics of
$\sigma(b, k)$. The ringing between them comes from Fourier transforming the
two sharp features in $k$ space into $l$ space. These dips and
ringing at structure scales are a result of the enhancement of power at these scales originating from the channel of the nonlocal entanglement. As
we discussed above, the first one near $l\simeq 180$ may be correlated
to the void appearing at the same scale, while the second damping
originates from the potential modification term switching from
negative to positive thereby going through a discontinuity at around
$k=40$, Fig.~\ref{fig:fig4}. These features may be distinguishable by future surveys.

\section{A Smoking Gun for The Landscape?}
\label{sec:smoke}

Much of the current discussion about  string theory is centered around the question of whether any theory with a landscape of vacua as well populated as that of string theory could ever be predictive. Perhaps all we can do is look for correlations between various observables in each of the vacua of the landscape and make a very restricted set of ``predictions'' from these correlations. This is a rather bleak view of the future of physics, and one we do not share.

What we have shown in this work as well as its prequel is that, while we freely admit that the detailed behavior of the string landscape is still unknown, it is not out of the realm of possibility that we might be able to detect ``smoking-gun'' effects from it on the sky. In particular, the effects of the dynamics of quantum gravity, and in particular, those due to nonlocal entanglements between our horizon patch and others, could modify the Friedmann equation and hence other cosmological observables and these modifications could be amenable to present or near future observations. 

Let us summarize our results in more detail.

In paper $I$ \cite{avatars1}  we showed that the requirement of having a
sufficiently flat inflationary potential {\it after} the modifications
to the Friedmann equation are taken into account, coupled with the
known value of the CMB quadrupole placed stringent bounds on the energy
scale related to the structure of vacua in the non-SUSY part of the
landscape. In our picture, this is the SUSY breaking scale, and we find
that it has to be significantly larger (five to eight orders of
magnitude larger) than studies of the hierarchy problem would have
required. The LHC will be able to test this statement soon, once again
showing the tight interconnectivity between particle physics and
cosmology.  

Besides the tight bound on SUSY breaking scale, we find
many interesting and perhaps unique clues imprinted on the CMB
and on large scale structure.

We have shown here how the negative, scale-dependent, energy
correction in the modified Friedman equation suppresses the
temperature anisotropy spectrum at large angles, Fig.~\ref{fig:fig1}
which could help explain the observed suppression of power in the
$TT$-spectrum by WMAP and COBE . The second and the third peak seem to
also be mildly sensitive to the SUSY breaking scale $b$ and to the scale of
the nonlocal entanglement $L_1$. As we have explained in the text this
is related to the enhancement of power Fig.~\ref{fig:fig2} at structure
and substructure scales and the scale-dependent modifications to the
background Newtonian potential given by Eq.~(\ref{eq:modepot}) which
act as a source for the $TT-$ spectrum.

The matter power spectrum is given in Fig.~\ref{fig:fig2} and the
modulation at scales after the last scattering shows up there too,
although it is a small effect. However, as can be seen from
Fig.~\ref{fig:fig2}, there are definite imprints on $P(k)$ with their
origin in the corrections to the Newtonian potential coming from the
quantum gravitational entanglement. The correction to the background
Newtonian potential coming from the entanglement of our universe with
superhorizon fluctuations of our inflaton patch and others leaves
a definite imprint by damping structure at large scales and
enhancing it at structure and substructure scale. 
As far as the linear theory is concerned, 
these corrections bring the matter power spectrum into
much better agreement than LCDM with the data from the recent result of 
SDSS \cite{SDSS}. This also suggests that it may have implications 
for nonlinear modeling.

The effects we find here also provide a natural mechanism for creating
primordial voids only at relevant scales $k\simeq 1$ and $k \simeq
20$. These effects come from the potential correction of the
non-Gaussian nonadiabatic superhorizon fluctuations being superimposed
on the background potential of the gaussian perturbations thereby
giving rise to a negative density contrast $\delta_f <0$ at certain
scales discussed above. The implications of this superimposition are
twofold: It induces a running in the scalar spectral index $n_s$ as
calculated above, $n_s=n^0_{s} +\delta n_s$, which is in agreement
with the WMAP data  \cite{wmap3} and; it creates voids at large
scales where the background and correction terms are comparable.

We showed in paper $I$ \cite{avatars1} that the anisotropy scale contributing
to the spectrum is bound to be at least 100 times greater than the
horizon radius namely $10^{-10}\ M_P <b<10^{-8}\ M_P$, corresponding
to roughly the first $1-10$ e-foldings for GUT inflation. In terms of
energies this stretches to about 18 e-foldings. This corresponds to
primordial voids being created at $k\simeq 1$ or $50-60$ degrees,
(causally disconnected part), in the sky or roughly at
$l<10$. Following the calculation done by Inoue and Silk \cite{inouesilk}
at the present epoch, the angle $\theta$ is given by the ratio of the
size $r_v$ to the distance $d$: $\sin\theta \simeq w = a r_v \slash
d$. Their estimate shows that for voids with $w\simeq 0.9, r_v \simeq
r_H \slash10$ they agree with the SDSS findings of the observed two
cold spots in the sky. As we have shown here the first voids are
formed at $k\simeq1, \theta \simeq 60$ which is in agreement with the
SDSS findings.Thus the void induced quadruple suppression may explain
the two observed cold spots since $\Delta T / T \simeq
\delta\Phi/\Phi$

The anticorrelation between the negative and positive density
contrasts at low $l$  ($l<10$) shows itself as an alignment between the quadruple/octupole with more
power concetrated along this direction \cite{landmagueijo,tegmark} as
shown in \cite{inouesilk} {\it i.e.} a preferred direction along the correlation
where the angular momentum dispersion is a maximum. The preferred
direction is the line parallel to the vector connecting the centers of
the voids located at 60 degrees. The superhorizon potential
fluctuation related to the quadrupole scale acts as a gradient field
with the quadrupole direction as the preferred axis. A more careful
treatment of this issue requires a more detailed calculation of the
evolution of the negative density contrasts, especially for the other
voids located at $k=20$, as well as the generation of the sky maps for
this model. This is beyond the scope of this paper.

Our smoking gun which uniquely identifies the CMB and LSS signatures of this model, comes from
cross-correlating cosmic shear with temperature anisotropies. As
discussed in \cite{steenlaura} this cross-correlation compares the
sources that seed the lensing potential which maps large scale
structure and the source that seed CMB anisotropies. In the case of
conventional inflation that source shoud be the primordial spectrum
$P_{\mathcal R}$ and be the same for both CMB and LSS. Therefore a
conventional cosmology predicts a correlation of order one. What we
find here instead is that only the correlation is not close to one ,
see Fig.~\ref{fig:fig4} thereby demonstrating the existence of
noninflationary channels contributing to structure, but that there are
unique and rather interesting features in this spectrum at
structure and substructure scales which soon will be observed by the 
PLANCK mission. This unique features come from the highly notrivial expression
of the nonlocal entanglement contained in $F(b,V), L_1$. We can describe
it physically by noting that the discontinuities induced on the scalar
spectral index $n_s$ show as the dips in Fig.~\ref{fig:fig4}. when
Foorier transformed from k space to l-space, these together with the
scale-dependent corrections on the Newtonian potential $\delta_f =
\delta\Phi / \Phi^0$ show as ringing in the correlation for scales
shorter after the surface of the last scattering. These are clearly
very definite signatures that can be looked for by future surveys.

The main lesson to be taken from this, admittedly speculative,
treatment of the dynamics of the landscape is that there are
concrete calculations that can be done and concrete predictions that
can be made. We certainly do not argue that our results are {\em exactly} how the landscape behaves. Rather, there are interesting possibilities that may be robust enough to survive a more detailed
analysis of the landscape. When a more detailed theory
of the quantum gravity in the early universe is known, as we have
demonstrated here, its astrophysical imprints may be within reach of
observation in the cosmological arena.

\begin{acknowledgments}

L.~M-H would like to thank K.~Land and J.~Magueijo for their help with
voids. L.M-H was supported in part by DOE grant
DE-FG02-06ER1418 and NSF grant PHY-0553312. TT thanks K.~T.~ Inoue for useful discussion.  R.~H. was supported in part by DOE grant DE-FG03-91-ER40682. He would also like to thank the Perimeter Institute for their generous hospitality while
this work was in progress.
\end{acknowledgments}

\end{document}